\documentclass[twocolumn,aps,amsmath,amssymb,prl,showpacs]{revtex4}

\usepackage{graphicx}
\usepackage{bm}
\usepackage{times}
\usepackage{color}

\usepackage{graphicx,color,rotating}
\newlength{\plotwidth}
\newcommand{\im}{\mathrm{i}}

\newcommand{\ket}[1]{\,\big|{#1}\big> }

\newcommand{\matrixe}[3]{\big<{#1}\big|\,{#2}\big|\,{#3}\big> }

\newcommand{\op}[1]{\hat{\mathrm{#1}}}

\newcommand{\conop}[1]{\op{#1}^{\dagger}}
\newcommand{\desop}[1]{\op{#1}^{\phantom{\dagger}}}

\newcommand{\OH}{\op{H}}

\begin{document}
\title{Ultracold Bose gases in time-dependent 1D superlattices: response and quasimomentum structure}

\author{Markus Hild}
\email{markus.hild@physik.tu-darmstadt.de}
\author{Felix Schmitt}
\author{Ilona T\"urschmann}
\author{Robert Roth}
\affiliation{Institut f\"ur Kernphysik, Technische Universit\"at Darmstadt, Schlossgartenstr. 9, 64289 Darmstadt, Germany}

\date{\today}

\begin{abstract}
The response of ultracold atomic Bose gases in time-dependent optical lattices is discussed based on direct simulations of the time-evolution of the many-body state in the framework of the Bose-Hubbard model. We focus on small-amplitude modulations of the lattice potential as implemented in several recent experiment and study different observables in the region of the first resonance in the Mott-insulator phase. In addition to the energy transfer we investigate the quasimomentum structure of the system which is accessible via the matter-wave interference pattern after a prompt release. We identify characteristic correlations between the excitation frequency and the quasimomentum distribution and study their structure in the presence of a superlattice potential.  
\end{abstract}

\maketitle


Ultracold atomic gases in optical lattices are a versatile laboratory for the study of fundamental quantum phenomena. The accurate control of the important physical parameters over a wide range has been utilized for detailed experimental investigations of quantum phase transitions, e.g. the superfluid to Mott insulator phase transition (SF-MI) \cite{JaBr98,GrMa02a,GrMa02b}. The primary experimental observable is the matter-wave interference pattern of the atoms after release from the confining potentials and ballistic expansion. The interference pattern obtained after a sudden release of the atoms provides direct information on the (quasi)momentum distribution of the system before the release. The interference pattern thus helps with the identification of different quantum phases, such as the SF and MI phases \cite{GrMa02a,GrMa02b}. In recent experiments it was also used to study the response of the system in a two-photon Bragg-spectroscopy scheme based on a temporal modulation of the lattice potential. The broadening of the central interference peak after re-thermalization in a shallow lattice was used as a measure for the response \cite{StMo04,FaLy07}. These techniques can also be employed in the context of more complicated irregular lattices as they can be produced, e.g., by using speckle patterns \cite{LyFa05,ClVa05,FoFa05,ScDr05} or two-color superlattice potentials \cite{RoBu03b,RoBu03c,FaLy07}. In pioneering experiments on the response in two-color superlattices the impact of a superlattice has been investigated \cite{FaLy07}. In agreement with theoretical predictions a broadening of the response as function of modulation frequency was observed with increasing superlattice amplitude.  

Motivated by these experiments we study the response of a bosonic system to the modulation of the superlattice potential deep in the Mott-insulating regime. Going beyond the observables used in experiments, we study the quasimomentum structure of the system as it is revealed by the interference pattern after a prompt release of the atoms without any re-thermalization phase. To this end we perform an explicit time-evolution of the many-body state in the presence of a time-dependent lattice based on the Bose-Hubbard Hamiltonian \cite{HiSc06}. Similar studies have been done using the time-dependent density matrix renormalization group method (tDMRG) \cite{KoIu06}. Our simulations reveal subtle correlations between the modulation frequency within the resonance region and the quasimomentum distribution, which should be accessible to experiment.


A system of $N$ bosons in an one-dimensional superlattice potential with $I$ sites at zero temperature is well described by the single-band Bose-Hubbard model (BHM) \cite{JaBr98}. The Bose-Hubbard Hamiltonian, formulated in second quantization in the basis of localized Wannier states of the lowest bands, reads
\begin{align}\label{eq_sec2_hamiltonian}
 \OH_0=	 -J_0\!\sum_{\langle{}i,j\rangle}\conop{a}_i\desop{a}_j
		 +\frac{U_0}{2}\!\sum_{i}\op{n}_i(\op{n}_i\!-\!1)
		 +\Delta_0\!\!\sum_{i}\epsilon_i\op{n}_i,
\end{align}
where the first sum runs over adjacent sites including a term connecting the first and last site of the lattice (cyclic boundary conditions). The Hamiltonian \eqref{eq_sec2_hamiltonian} with the bosonic creation (annihilation) operators $\conop{a}_i$~($\desop{a}_i$) and the occupation number operator $\op{n}_i$ consists of three terms: The first term describes the tunneling between adjacent sites, the second term accounts for the on-site interaction of the atoms, and the third term represents a site-dependent external potential. We introduce the superlattice potential via the latter term and describe its spatial structure by the reduced on-site energies $\epsilon_i \in [-1,0]$. The physics of the BHM is governed by the competition between the relative strengths of these three terms, i.e. the tunneling strength $J_0$, the interaction strength $U_0$, and the superlattice amplitude $\Delta_0$. 

The basis of the BHM for $N$ bosons and $I$ sites is spanned by the occupation number states $\ket{\{n_1,n_2,\dots,n_I\}_{\alpha}}$ for all possible sets of occupation numbers $n_i$ with $\sum_{i} n_i = N$. An arbitrary state can be expanded in this number-state basis leading to 
$\ket{\psi}=\sum_{\alpha=1}^D C_{\alpha}\ket{\{n_1,n_2,\dots,n_I\}_{\alpha}}\label{eq_sec2_state}$
with complex coefficients $C_{\alpha}$. The coefficients of the eigenstates $\ket{\nu}$ can be obtained by the numerical solution of the eigenvalue problem of the Hamilton matrix.

\begin{figure*}[!t]
\centering\includegraphics[width=0.8\textwidth]{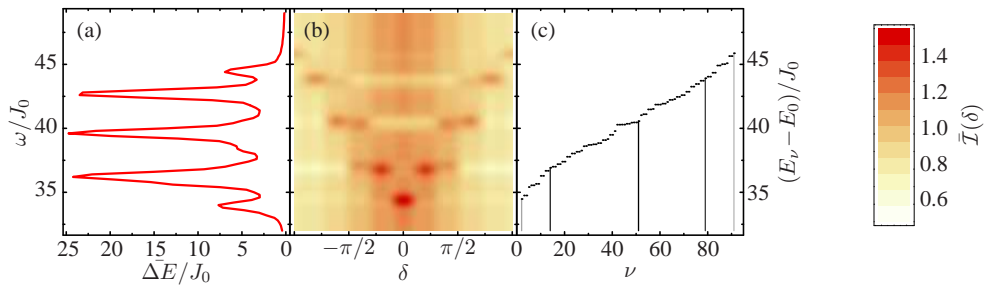}
\caption{(color online) The 1U-resonance in a system with $I\!=N\!=10$ at $U_0/J_0\!=\!40$ for a regular lattice ($\Delta_0\!=\!0$). Panel (a) illustrates the time-averaged energy transfer $\bar{\Delta E}$ versus modulation frequency $\omega$ of the lattice. Panel (b) shows the corresponding time-averaged interference pattern which also represents the quasimomentum distribution. Panel (c) shows the energy spectrum in the region of the first Hubbard-band with the vertical lines indicating sizable matrix elements $\matrixe{0}{\OH_J}{\nu}$ (darker shadings refer to larger values). The truncation energy of the basis is $E_\text{trunc}/J_0\!=\!120$.}
\label{fig:etrans_int}
\end{figure*}

%
The basis dimension $D$ grows factorially with $I$ and $N$, thus limiting any calculation using the full basis to small systems. However, in the strongly interacting regime ($U_0\gg{}J_0$) only a few basis states contribute to the low-lying eigenstates. This allows for a physically motivated truncation \cite{HiSc06,ScHi07} of the many-body basis. The relevant basis states are identified using the expectation value of the Hamiltonian $\matrixe{\{n_1,n_2,\dots,n_I\}}{\OH_0}{\{n_1,n_2,\dots,n_I\}} \leq E_\text{trunc}$, and only states below the truncation energy $E_\text{trunc}$ are included. By varying $E_\text{trunc}$ one can explicitly assess and control the impact of the truncation on observables. For regular lattices ($\Delta_0\!=\!0$) and filling factor $N/I\!=\!1$ this truncation allows for all relevant $n$-particle--$n$-hole states with $n \leq E_\text{trunc}/U_0$ with respect to the reference state $\ket{1,1,\dots,1}$. 

%
We investigate the dynamics and response of the system induced by external time-dependent perturbations based on the explicit time evolution of the many-body state. 
Our calculations are motivated by recent experiments \cite{StMo04,FaLy07} using a sinusoidal modulation of the lattice depth to perform two-photon Bragg spectroscopy. Unlike our calculations, these experiments include a rethermalization phase in the superfluid regime before the time-of-flight measurement in order to assess the energy transfer to the system. We assume a prompt release without rethermalization to directly access the quasimomentum distribution after a certain modulation time.

Formally, the time-dependent lattice potential is written as $V(x,t)=[1+F\sin(\omega t)]\,V(x)$, where $V(x)$ is the static spatial lattice, $\omega$ the frequency, and $F$ the relative amplitude of the modulation. All simulations are performed with a small amplitude $F=0.1$ in accord with experiment. The time-dependence enters the Bose-Hubbard Hamiltonian $\OH(t)$ via the time-dependent parameters $J(t)$, $U(t)$, and $\Delta(t)$, which are obtained within a Gaussian approximation for the localized Wannier functions \cite{KBM05,HiSc06}. The parameters oscillate around their initial values $J_0$, $U_0$, and $\Delta_0$ at $t\!=\!0$. 

We investigate the response of the system deep inside the Mott-insulating regime for fixed $U_0/J_0=40$. The modulation frequency $\omega$ is varied in the range $\omega/J_0=32$ to $52$, which covers the so-called 1U resonance at $\omega/J_0\approx U_0/J_0 = 40$. The ground state obtained for the initial Bose-Hubbard Hamiltonian $\OH_0$ is used as initial state $\ket{\psi,0}$ and evolved up to $t J_0=10$ for each frequency $\omega$. The exact time-evolution is performed using a Crank-Nicholson scheme \cite{ScGa00,ScGa04} with typically $3000$ time steps. The relevant observables are evaluated at every 30th step. Following an initial phase with large changes, the observables saturate \cite{KoIu06} and show only minor fluctuations \cite{HiSc06}. Since this residual time-dependence will not be resolved in experiment, we average the observables over an interval of evolution times within the saturated regime from $t J_0=6$ to $10$ (time-averaged quantities indicated by a bar).

The simplest theoretical quantity to characterize the response of the system to the lattice modulation is the energy transfer $\Delta E(t)\!=\!\matrixe{\psi,t}{\!\OH_0}{\psi,t}\!-\!\matrixe{\psi,0}{\!\OH_0}{\psi,0}$. As an example, Fig. \ref{fig:etrans_int}(a) shows the energy transfer for a system with $I=N=10$ in the Mott-insulator phase ($U_0/J_0=40$) for a regular optical lattice ($\Delta_0=0$) in the vicinity of the 1U resonance at the modulation frequency $\omega=U_0$. A detailed analysis of the energy transfer is given in Ref. \cite{HiSc06}. 

%
Additional information can be obtained by a linear response analysis \cite{ClJa06,IuCa06,HiSc06}. To this end, the Hamiltonian is linearized with respect to the modulation amplitude $F$, leading to 
$\OH_{\text{lin}}(t) = \OH_0+FV_0\sin(\omega t) [\mu\OH_0-\kappa\OH_J ]$
with the amplitude $V_0$ of the lattice potential and the couplings $\mu$ and $\kappa$. The first part of the linear term is irrelevant, because it only generates an energy shift. The second part couples the ground state to excited states via the tunneling operator $\OH_J=-J\sum_{\langle i,j\rangle}\conop{a}_i\desop{a}_j$. In a linear response picture, resonant transitions from the ground state $\ket{0}$ are expected whenever the frequency $\omega$ coincides with the energy $E_{\nu}$ of an excited state $\ket{\nu}$ with a sizable matrix element $\matrixe{0}{\OH_J}{\nu}$. This interrelation is illustrated in Fig. \ref{fig:etrans_int}(c), where the energy eigenvalues for the first Hubbard band are shown. The vertical lines indicate excited states with sizable matrix elements $\matrixe{0}{\OH_J}{\nu}$. According to the linear response analysis, the energies of those eigenstates should correspond to the resonance frequencies in the response. The comparison of the resonance energies emerging in the energy transfer in Fig. \ref{fig:etrans_int}(a) with these excitation energies in \ref{fig:etrans_int}(c) confirms this interpretation.

%
In addition to the energy transfer we investigate the evolution of the interference pattern. The matter-wave interference pattern after ballistic expansion is used to gain experimental information on the many-body state. In recent experiments \cite{StMo04,FaLy07} the central peak of the interference pattern was employed as a measure for the response of the gas to a lattice modulation. The intensity distribution $\mathcal{I}(\delta)$ of the interference pattern as function of the relative phase $\delta$ is given by
$\mathcal{I}(\delta)=\frac{1}{I}\sum_{i,j}\exp[\im(i\!-\!j)\delta]\;\matrixe{\psi}{\conop{a}_i\desop{a}_j}{\psi}$ for an arbitrary state $\ket{\psi}$ \cite{RoBu03a}. For $\delta=\!2\pi q/I$ the intensity $\mathcal{I}(\delta)$ corresponds to the occupation numbers $n_q\!=\!\matrixe{\psi}{\conop{c}_q\desop{c}_q}{\psi}\!=\!\mathcal{I}(\delta\!=\!2\pi q/I)$
of quasimomentum eigenstates with $q\!=\!\delta \,I/2\pi$. Here, $\conop{c}_q$\,($\desop{c}_q$) are the bosonic creation (annihilation) operators with respect to the quasimomentum basis, which can be written as $\conop{c}_q=\frac{1}{\sqrt{I}}\sum_{i}\exp[\im\tfrac{2\pi}{I}qi]\;\conop{a}_i$.

Figure \ref{fig:etrans_int}(b) illustrates the frequency-dependence of the interference pattern in the region of the 1U resonance. Each horizontal cut through the density plot represents the time-averaged interference pattern for a certain frequency $\omega$. We assume that the lattice is switched off instantaneously after a certain evolution time in the modulated lattice---different from recent experiments which involve a re-thermalization period in the superfluid regime \cite{StMo04,FaLy07}. The general interference structure reveals a specific correlation between the frequency $\omega$ relative to the centroid of the 1U resonance and the quasimomentum distribution, i.e. the peaks of the interference pattern. Away from the resonance region, the intensity $\mathcal{I}(\delta)$ exhibits a broad background distribution characteristic for the Mott-insulating phase. For frequencies $\omega$ at the low-frequency end of the resonance a sharp interference peak emerges at $\delta=0$ indicating the resonant transition to the $q=0$ state.  With increasing frequency $\omega$ this population moves to successively higher quasimomenta $|q|$, i.e. the interference peak splits and shifts towards larger $|\delta|$. The fine-structure of the resonance is thus mapped onto the interference pattern in an experimentally accessible way.

\begin{figure}[t]
\includegraphics[width=0.95\columnwidth]{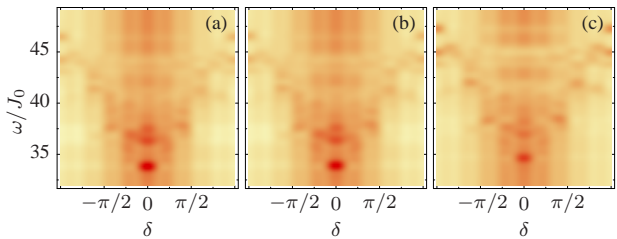}
\caption{(color online) Interference pattern as function of $\omega$ in the region of the 1U resonance for a system with $I\!=N\!=10$ at $U_0/J_0\!=\!40$ and $\Delta_0/J_0\!=\!2$ for truncation energies $E_\text{trunc}/J_0\!=\!120$ (a), $E_\text{trunc}/J_0\!=\!80$ (b), and $E_\text{trunc}/J_0\!=\!40$ (c). The color coding is the same as in Fig.\ref{fig:results_ic}.}
\label{fig:benchmark}
\end{figure}
\begin{figure}[t]
\includegraphics[width=0.8\columnwidth]{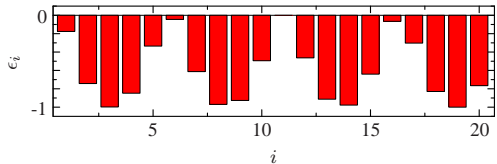}
\caption{Distribution of the $\epsilon_i$ for $I\!=\!20$ lattice sites of the incommensurate superlattice  used in the calculations.}
\label{fig:lattice_struct}
\end{figure}

%
So far, the simulations were restricted to small systems with large bases. To treat larger lattices we reduce the truncation energy $E_{\text{trunc}}$ controlling the basis size. Figure \ref{fig:benchmark} illustrates the insensitivity of the interference pattern on changes of $E_{\text{trunc}}$. There is practically no difference when reducing the truncation energy from $E_\text{trunc}/J_0=120$ to $80$. Even for $E_\text{trunc}/J_0=40$ all relevant features are reproduced, although the intensity of the interference peaks shows slight deviations. All qualitative conclusions regarding the correlations between frequency and quasimomentum distribution remain unaffected.

%
Using the lowest truncation energy we investigate the response of a system with $I\!=\!N\!=\!20$ at $U_0/\!J_0\!=\!40$. The major focus is on the change of the response and the interference pattern if a two-color superlattice potential of increasing amplitude $\Delta_0$ is added. The distribution of the relative strengths $\epsilon_i$ for the incommensurate superlattice used in the following are shown in Fig. \ref{fig:lattice_struct}. The the ratio of the wavelengths of the two standing waves is $\lambda_1/\lambda_2\!\approx\!0.81$, similar to a recent experiment \cite{FaLy07}.

\begin{figure}[t]
\includegraphics[width=0.95\columnwidth]{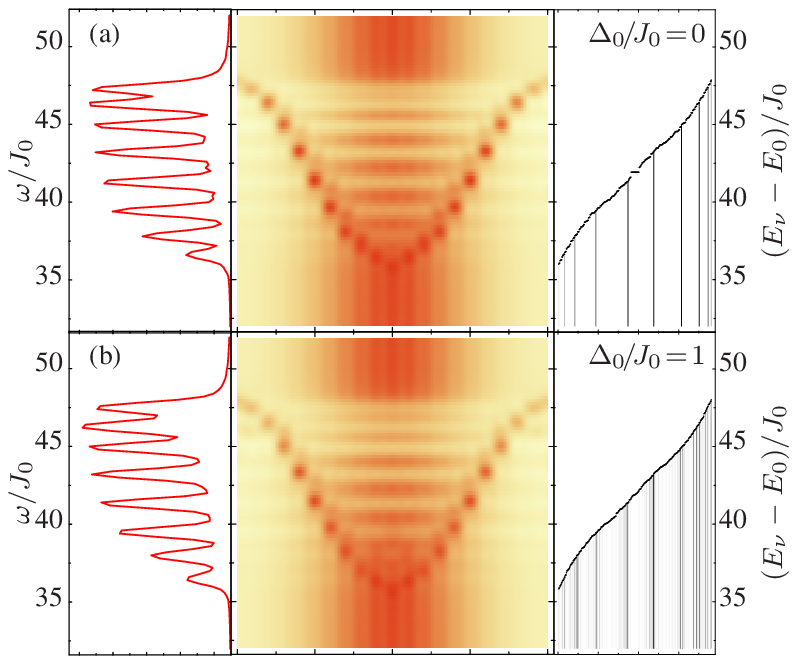}\\[-1.5pt]
\includegraphics[width=0.95\columnwidth]{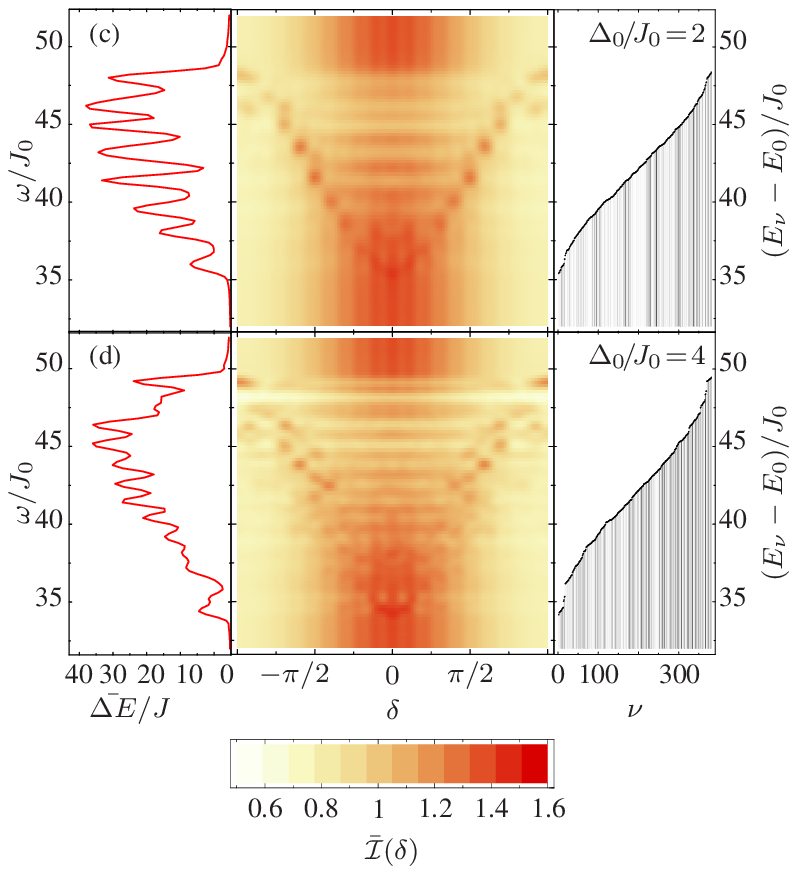}
\caption{(color online) Energy transfer (left-hand panels), interference pattern (middle panels), and excitation spectrum (right-hand panels) of a system with $I\!\!=\!\!N\!\!=\!\!20$ and $U_0/J_0\!\!=\!\!40$ for several superlattice amplitudes $\Delta_0$ as indicated in the plots. The density plots illustrate the correlations between quasimomentum distribution and excitation frequency in the region of the 1U-resonance. The vertical lines in the energy spectra (right-hand panels) point out strong matrix elements $\matrixe{0}{\OH_J}{\nu}$.}
\label{fig:results_ic}
\end{figure}

Figure \ref{fig:results_ic} depicts the evolution of the interference structure as function of the superlattice amplitude $\Delta_0$ for the incommensurate case. The right-hand panels show the energy spectrum with vertical lines marking sizeable matrix elements $\matrixe{0}{\OH_J}{\nu}$. The left-hand panels depict the energy transfer as function of frequency. The result for $\Delta_0/J_0\!=\!0$ shown in Fig. \ref{fig:results_ic}(a) confirms our previous discussion of the smaller regular lattice in Fig. \ref{fig:etrans_int}. The linear response analysis provides a good estimate for the resonance energies via the energies of those excited states that exhibit strong transition matrix elements to the ground state. Furthermore, the correlation between frequency $\omega$ and the quasimomentum distribution is even more pronounced. At the low-frequency end of the resonance quasimomentum states around $q\approx0$ are populated, whereas for larger modulation frequencies successively higher quasi-momenta are occupied. 

The results for a small superlattice amplitude $\Delta_0/J_0\!=\!1$ in Fig. \ref{fig:results_ic}(b) show minor changes in the quasimomentum structure and the energy transfer as compared to (a). Nevertheless, the number of possible excitations from the ground state increases. A further increase of the superlattice amplitude to $\Delta_0/J_0\!=\!2$ leads to a weak suppression of the interference structure as shown in Fig. \ref{fig:results_ic}(c). In comparison to the energy spectrum in (b) there are many more large matrix elements which are not localized at distinct energies but spread over the whole range. The occurrence of the small gaps at both ends of the energy band is also visible in the density plots as small shifts in the interference structure along the energy (vertical) axis. This also indicates the broadening of the resonance due to the superlattice \cite{FaLy07,HiSc06}. 

Further increase of the superlattice amplitude to $\Delta_0/J_0\!=\!4$ leads to the disappearance of the interference peaks as depicted in Fig. \ref{fig:results_ic}(d). The number of strong matrix elements which couple ground and excited states is further increased. Although the interference pattern does not show sharp peaks anymore, the energy transfer in these cases still exhibits a strong resonance behavior \cite{HiSc06}. Only the fine-structure of the resonance in the energy transfer is affected by the superlattice. However, the distinct correlations between excitation frequency and quasimomentum distribution vanish far below the transition from the homogeneous Mott insulator to the Bose-glass phase at $\Delta_0\approx{}U_0$. These general features are robust against changes of the superlattice structure. We have performed explicit simulations using a commensurate superlattice showing the same behavior. 

The rapid change of the interference pattern can be explained in the linear response picture: In the absence of a superlattice, resonant transitions connect the ground state to a few excited states only---those  characterized by large transition matrix elements $\matrixe{0}{\OH_J}{\nu}$ [Fig. 4(a) and (b)]. The many-body state $\ket{\psi,t}$ during the time evolution is dominated by these few states and exhibits well-defined interference peaks. With increasing superlattice amplitude $\Delta_0$ more and more sizable transition matrix elements emerge and the time-evolved state is a superposition of a large number of eigenstates [Fig. 4(c) and (d)]. The fragmentation of the state causes a fragmentation of the interference pattern and in effect a suppression of the distinct peaks. This mechanism  can be confirmed within a simple toy-model by comparing the interference pattern of an excited eigenstate with the one resulting from a coherent superposition of a few neighboring eigenstates. Note that this phenomenon is quite different from the Mott-insulator to Bose-glass transition, which appears at much larger superlattice amplitudes.

In summary, we have discussed the response of Bose gases in modulated lattice and superlattice potentials with an emphasize on the quasimomentum distribution and the interference pattern after prompt release from the lattice. For modulation frequencies in the region of the 1U resonance distinct peaks appear in the interference pattern. Their position is correlated with the modulation frequency relative to the centroid of the resonance: with increasing frequency they shift to larger quasimomenta. In the presence of a superlattice this characteristic correlation vanishes already for small superlattice amplitudes, much faster than the resonance observed in the energy transfer.


\begin{thebibliography}{10}
\expandafter\ifx\csname url\endcsname\relax
  \def\url#1{\texttt{#1}}\fi
\expandafter\ifx\csname urlprefix\endcsname\relax\def\urlprefix{URL }\fi

\bibitem{JaBr98}
D.~Jaksch \emph{et~al.}, Phys. Rev. Lett. \textbf{81} (1998) 3108.

\bibitem{GrMa02a}
M.~Greiner \emph{et~al.}, Nature \textbf{415} (2002) 39.

\bibitem{GrMa02b}
M.~Greiner \emph{et~al.}, Nature \textbf{419} (2002) 51.

\bibitem{StMo04}
T.~St\"oferle \emph{et~al.}, Phys. Rev. Lett. \textbf{92} (2004) 130403.

\bibitem{FaLy07}
   L.~Fallani \emph{et~al.}, Phys. Rev. Lett. \textbf{98} (2007) 130404.

\bibitem{LyFa05} 
  J. E. Lye \emph{et~al.}, Phys. Rev. Lett. \textbf{95} (2005) 070401.

\bibitem{ClVa05}
  D.~Clement \emph{et~al.}, Phys. Rev. Lett. \textbf{95} (2005) 170409.

\bibitem{FoFa05} 
  C. Fort \emph{et~al.}, Phys. Rev. Lett. \textbf{95} (2005) 170410.
  
\bibitem{ScDr05}
  T.~Schulte \emph{et~al.}, Phys. Rev. Lett. \textbf{95} (2005) 170411.

\bibitem{RoBu03b}
   R.~Roth, K.~Burnett, J. Opt. B: Quantum Semiclass. Opt. \textbf{5} (2003) S50.

\bibitem{RoBu03c}
   R.~Roth, K.~Burnett, Phys. Rev. A \textbf{68} (2003) 023604.

\bibitem{HiSc06}
  M.~Hild, F.~Schmitt, R.~Roth, J. Phys. B \textbf{39} (2006) 4547.

\bibitem{KoIu06}
  C.~Kollath \emph{et~al.}, Phys. Rev. Lett. \textbf{97} (2006) 050402.

\bibitem{ScHi07}
  F.~Schmitt, M.~Hild, R.~Roth, J. Phys. B \textbf{40} (2007) 371.

\bibitem{KBM05}
  K. Braun-Munzinger, Ph.D. thesis; Oxford (2005).

\bibitem{ScGa00}
  O.~Schenk \emph{et al.}, BIT \textbf{40} (2000) 158.

\bibitem{ScGa04}
  O.~Schenk, K.~G\"artner, FGCS \textbf{20} (2004) 475.

\bibitem{ClJa06}
  S.~R.~Clark, D.~Jaksch, New J. Phys. \textbf{8} (2006) 160.

\bibitem{IuCa06}
  A.~Iucci \emph{et~al.}, Phys. Rev. A \textbf{73} (2006) 041608(R).

\bibitem{RoBu03a}
  R.~Roth, K.~Burnett, Phys. Rev. A \textbf{67} (2003) 031602(R).

\end{thebibliography}
\end{document}